\documentclass[twocolumn,apl,sd,amsmath,amssymb,reprint]{revtex4-1}

\usepackage{graphicx}% Include figure files
\usepackage{dcolumn}% Align table columns on decimal point
\usepackage{bm}% bold math
\usepackage{xcolor} % black, blue, brown, cyan, darkgray, gray, green, lightgray, lime, magenta, olive, orange, pink, purple, red, teal, violet, white, yellow.

\graphicspath{}
\begin{document}
\title[]{Surface acoustic wave unidirectional transducers for quantum applications}

\author{Maria K. \surname{Ekstr\"om}}\author{Thomas \surname{Aref}} \author{Johan \surname{Runeson}} \author{Johan \surname{Bj\"orck}} \author{Isac \surname{Bostr\"om}}\author{Per \surname{Delsing}}
% \email{per.delsing@chalmers.se}
 \affiliation{Chalmers University of Technology, Microtechnology and Nanoscience, 412 96 G\"oteborg}
\date{\today}

\begin{abstract}
The conversion efficiency of electric microwave signals into surface acoustic waves in different types of superconducting transducers is studied with the aim of quantum applications. We compare delay lines containing either conventional symmetric transducers (IDTs) or unidirectional transducers (UDTs) at 2.3~GHz and 10~mK.
The UDT delay lines improve the insertion loss with 4.7~dB and a directivity of 22~dB is found for each UDT, indicating that 99.4~\% of the acoustic power goes in the desired direction. The power lost in the undesired direction accounts for more than 90~\% of the total loss in IDT delay lines, but only $\sim$3~\% percent of the total loss in the FEUDT delay lines.
\end{abstract}

\keywords{surface acoustic wave, SAW, interdigital transducer, IDT, floating electrode unidirectional transducer, FEUDT, UDT gigahertz frequency, low temperature, cryogenic temperature, directivity, insertion loss, propagation loss, phonon, delay line, SAW filter, efficient conversion, quantum acoustics, superconducting circuits}
\maketitle

Surface acoustic waves (SAWs) are Rayleigh waves propagating on the surface of a solid \cite{Rayleigh1885}. It has recently been suggested \cite{MVG2012} and shown \cite{MVG} that SAWs can interact with artificial atoms at the quantum level. This is fundamentally interesting because the artificial atoms can be made much larger than the wavelength of the SAW, which is not possible in other systems \cite{Kockum2014}. There are extensive new possibilities for quantum devices utilizing SAW; such as resonators \cite{Manenti2016,Magnusson2015}, absorption in double quantum dots \cite{Naber2006SAWDQD}, transport of quantum information \cite{Barnes2000,Hermelin2011,McNeil2011} and phonon assisted tunneling \cite{Santos2010}. 

When SAWs are used to carry quantum information, it is important to have low losses. The purpose here is to lower the conversion loss between electric signals (photons) and SAWs (phonons). In all studies about quantum SAW applications, SAWs are converted to and from electric microwave signals using conventional symmetric interdigital transducers (IDTs). The IDT can be described by a three port scattering matrix, where port 1 and 2 are acoustic and port 3 is electric \cite{bookchapter}. It has the same electric to SAW conversion in both ports, \textit{i.e.} S$_{13}=$~S$_{23}$, and hence 50~\% of the power is converted in the wrong direction. This means that IDTs are limited by a theoretical minimum insertion loss of \mbox{-3~dB} and because of reciprocity delay lines with two IDTs are theoretically limited to \mbox{-6~dB}.

Unlike the symmetric IDT, a unidirectional transducer (UDT) \cite{Collins1969,Morgan} can be optimized to release most of its SAW energy in one preferred direction, by maximizing the scattering element S$_{13}$ while minimizing S$_{23}$. In this way UDTs can exceed the \mbox{-3~dB} loss, and therefore UDTs are interesting to study for quantum SAW applications. 

UDTs have previously been studied for classical applications, such as low-loss-SAW filters at room temperature \cite{Datta, Morgan}. Since they have complicated structures, a substantial effort has been made in engineering low loss UDTs at gigahertz frequencies \cite{Yamanouchi1992GHzFEUDT}. 
Various types of UDTs and combinations of piezoelectric materials have also been explored, and some experiments have utilized higher harmonics \cite{Yamanouchi1994highGHzFEUDT}. Although, the UDTs have been studied at gigahertz frequencies, there have been very few studies at low temperatures using superconduting transducers \cite{Yamanouchi1998scNb}. Superconducting transducers do not suffer from the resistive losses that limit their performance at room temperature. Here, we study superconducting UDTs and IDTs at 10~mK in order to use them for quantum applications as efficient electric/SAW converters. We also make a detailed analysis of the remaining losses. 

\begin{figure}
\includegraphics{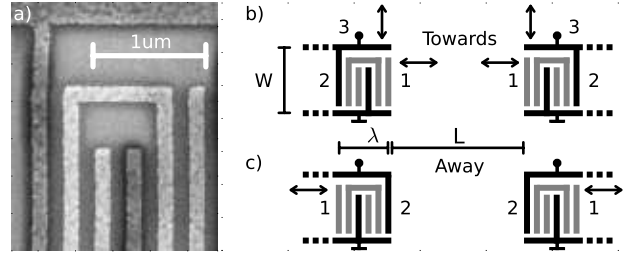}
\caption{\label{fig:FEUDT} a) Electron micrograph of the top part of one unit cell of a floating electrode unidirectional transducer. Note that the floating electrodes are brighter due to charging effects. The preferred electric/SAW conversion is towards the right \cite{Yamanouchi1984}, in port 1. b) A Towards delay line, where port 1 faces inwards and c) an Away delay line, where port 1 faces out from the delay line. One unit cell is illustrated, where the upper bus is connected to a live electrode, the lower bus is grounded and two of the floating electrodes (gray) are connected to each other.  }
\end{figure}

%\section{Design}
Both the UDTs and IDTs were designed for 2.3~GHz on lithium niobate. The UDT structure was selected from preliminary measurements of various types of UDTs at a lower frequency and room temperature. It is based on a floating electrode unidirectional transducer (FEUDT) \cite{Yamanouchi1984}, seen in Fig.~\ref{fig:FEUDT}a. As most types of UDTs it consists of a periodic structure, where the centers of transduction and reflection are separated in each unit cell, and hence each unit cell shows some directivity. Each unit cell consists of six electrodes, all with the same width: one live electrode connected to the upper bus, one grounded electrode connected to the lower bus and four floating electrodes, two of which are connected to each other. The design is such that electric/SAW conversion is optimized for port 1 and minimized for port 2 (Fig.~\ref{fig:FEUDT}b).

When conducting experiments at the quantum level, it is necessary to cool the system to cryogenic temperatures and operate at microwave frequencies where $k_BT\ll \hbar \omega$. This is well satisfied for 2.3~GHz and 10~mK. 

%\section{Fabrication}
At this frequency, the FEUDT electrodes have a finger width of $\lambda/12=125$~nm. The electrode structure was made using electron beam lithography and a liftoff process with 27~nm thick aluminum capped by 3~nm of palladium. They were connected to ground planes (5/85/10~nm of Ti/Au/Pd) in a delay line geometry. The transducers were separated edge to edge by 500~$\mu$m on the piezoelectric substrate YZ black lithium niobate (LiNbO$_3$). LiNbO$_3$ has strong piezoelectric coupling and is therefore especially interesting for quantum applications. 

An optimized device should have perfect impedance matching to 50~$\Omega$ and a maximum directivity. To obtain this, we varied both the electrode overlap $W$ (aperture of the transducer) and the number of unit cells $N_p$. The design parameters of the tested samples are summarized in Table \ref{tab:samples}. Two different electrode overlaps (W=35 and 46~$\mu$m) were investigated for both types of samples. The number of unit cells in the FEUDTs were either 110 or 160. For comparison, double electrode IDTs were designed to impedance match with 36 unit cells. The double electrode IDT structure minimizes internal reflections that otherwise complicate the IDT response \cite{Bristol1972,Morgan,Datta}.
\begin{table*}
\caption{\label{tab:samples}The maximum transmission~(\textit{Max T}) at frequency~($f_T$) and bandwidth for the Towards and the IDT delay lines. Each Towards delay line was compared with an Away delay line of the same type to retrieve the maximum transmission difference~(\textit{Max D}) at the corresponding frequency~($f_D$) and the frequency span where the difference was bigger than 20~dB. The total loss~($\gamma_\text{tot}$) was estimated from the directive loss~($\gamma_\text{D}$) of the transducers, and from loss due to viscous damping~($\gamma_\text{vis}$), beam steering~($\gamma_\text{bs}$) and diffraction~($\gamma_\text{diff}$) over the propagation distance $L+N_p\lambda$. $\gamma_\text{ue}$ is the loss that cannot be explained by directivity and propagation loss.}
\begin{ruledtabular}
\begin{tabular}{c c c | c c c  | c c c | c | c c c c | c | c}
\multicolumn{3}{c|}{Delay lines} &\textit{Max T}&$f_{T}$&\textit{BW}&\textit{Max D}&$f_{D}$&$>20~dB$&$L+N_p\lambda$&\multicolumn{5}{c|}{Estimated loss [dB]}&\\
Type& $N_p$ & $W$[$\mu$m]&[dB]&[GHz]&[MHz]&[dB]&[GHz]&[MHz]&[$\mu$m]&$\gamma_\text{D}$&$\gamma_\text{vis}$&$\gamma_\text{bs}$&$\gamma_\text{diff}$&$\gamma_\text{tot}$&$\gamma_\text{ue}$\\\hline
FEUDT\_1&110&35&-4.2&2.308&20&42&2.313&11&665&-0.06&-0.17&-0.37&-0.61&-1.2&-3.0\\
FEUDT\_2&110&46&-3.7&2.309&17&44&2.315&11&665&-0.06&-0.17&-0.28&-0.77&-1.3&-2.4\\
FEUDT\_3&160&46&-3.2&2.310&9.5&44&2.319&10&740&-0.06&-0.19&-0.31&-0.77&-1.3&-1.9\\
FEUDT\_4&160&46&-2.8&2.310&9.4&44&2.319&10&740&-0.06&-0.19&-0.31&-0.77&-1.3&-1.5\\\hline
IDT\_1&36&35		&-7.8&2.291&51&-&-&- 			&554&-6&-0.14&-0.31&-0.61&-7.1&-0.7\\
IDT\_2&36&35		&-7.7&2.290&51&-&-&- 			&554&-6&-0.14&-0.31&-0.61&-7.1&-0.6\\
IDT\_3&36&46		&-7.5&2.294&54&-&-&- 			&554&-6&-0.14&-0.23&-0.77&-7.1&-0.4\\
\end{tabular}
\end{ruledtabular}
\end{table*}

%\section{Measurements}
All the samples measured were delay lines. Four delay lines had FEUDTs with port~1 towards each other (Fig.~\ref{fig:FEUDT}b). They were optimized for high transmission and are described as "Towards". FEUDTs with the same design, but with port~1 facing out from the delay line are called "Away" (Fig.~\ref{fig:FEUDT}c). They were used for measurements of the SAW conversion via port~2, which is optimally very small. In this way, the power conversion via port~1 and port~2 could be compared (Row~1-4 in Table~\ref{tab:samples}). As a reference, three delay lines with double electrode IDTs were measured (Row~5-7 in Table~\ref{tab:samples}). 

\begin{figure}
\includegraphics{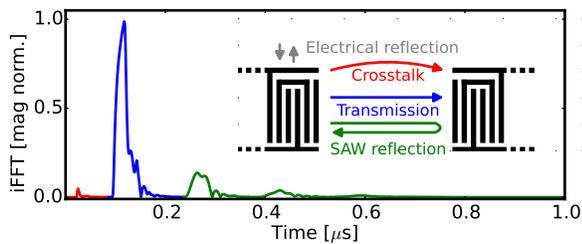}
\caption{\label{fig:FFT} Time domain data from Fourier transforming the transmission through FEUDT\_2. Towards. Separate peaks can be filtered selectively, where the first (red) is crosstalk, the second (blue) is the main SAW transmission and the remaining peaks (green) are SAWs transiting multiple times. (Inset:) FEUDT delay line. }
\end{figure}

%\section{Results}
Time domain data from Fourier transforming the measured signal through the delay lines showed separate peaks (Fig. \ref{fig:FFT}). The first peak was the electric crosstalk, the second the main transmission and the remaining peaks were SAWs transiting three or more times in the delay line. By selecting data only from the second (blue) peak, the main transmission could be isolated.

\begin{figure*}
\includegraphics{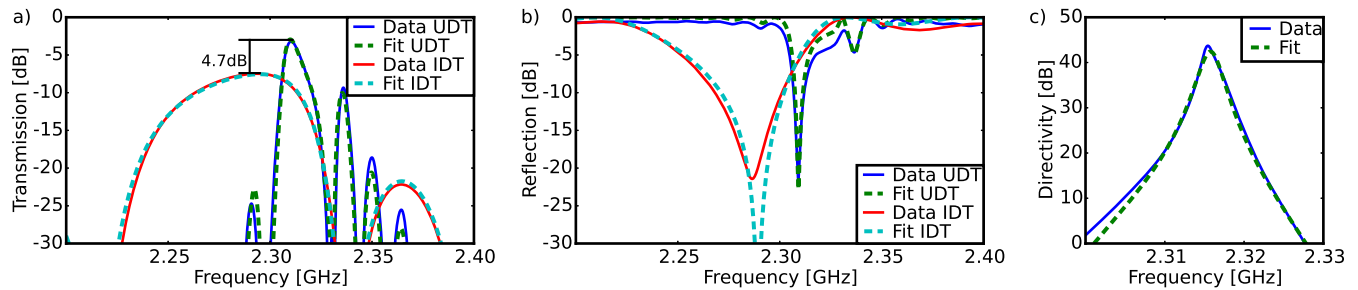}
\caption{\label{fig:RTD}
Response of FEUDT\_3 Towards (blue solid line) and IDT\_3  (red solid line). a) Transmission and b) reflection of the delay lines together with fits using the SAW circuit model for the IDT (cyan dotted line) and COM-theory for the FEUDT (green dotted lines). The discrepancy at minimum IDT reflection can be attributed to mechanical reflections and nonideal circulators that are not included in the circuit model, c) The difference between transmission through FEUDT\_2 Towards and Away (blue solid line) agrees with the COM-theory (green dotted lines).}
\end{figure*}

%\subsection{Fits to IDT circuit model and COM}
The reflection and transmission from the two delay lines with conventional IDTs matched well with the simplest SAW circuit model \cite{Smith1969,Morgan,Datta,bookchapter}, as seen in Fig.~\ref{fig:RTD}a,b. In this model, a double electrode IDT is described by an acoustic conductance $G_a(f)=G_{a0}\text{sinc}^2(\pi N_p(f-f_0)/f_0)$, an acoustic susceptance $B_a(f)$ (the Hilbert transformation of $G_a$) and a capacitance $C_T=\sqrt{2}N_pW\epsilon_{\infty}$.  $G_{a0}=4\cdot c_g^2 2\pi f_0 \epsilon_{\infty}N_p^2WK^2$ and is the acoustic conductance at center frequency $f_0$. $f$ is the driving frequency and $c_g\approx0.62$ \cite{Morgan} is a unitless factor accounting for the geometry of the electrodes. The room temperature literature values for the piezoelectric coupling coefficient ($K^2=4.8~\%$) and the effective permeability ($\epsilon_{\infty}=46\epsilon_0$) of LiNbO$_3$ were used in the fits. The only free fitting parameters were the center frequency and the attenuation.

Modeling the FEUDTs required a more complicated approach, which included internal mechanical reflections. The FEUDT results were in excellent agreement with the Coupled Mode (COM) theory \cite{Takeuchi1993,Morgan1999,Morgan2000,Morgan2001,Takeuchi1988}. The fitting parameters were the center frequency, the attenuation and $c_g$. 

Fig.~\ref{fig:RTD}a,b shows transmission and reflection for two different delay lines; a Towards FEUDT and an IDT delay line together with their fits. The transmission fits for both the IDT and the FEUDT were complemented with an attenuation factor to describe losses in the delay lines. The variation of the fitted $f_0$ corresponded to a few nanometers bigger electrode periodicity, which is within the range of error caused by the temperature difference and fabrication imperfections. 

%\subsection{Transmission maximum}
All Towards delay lines exceeded the maximum transmission of the IDT delay lines, see Table~\ref{tab:samples} and Fig.~\ref{fig:RTD}a. For the FEUDT delay lines with 160 unit cells, the maximum transmission was on average 4.7~dB higher than for the IDT delay lines. 

%\subsection{Directivity}
Comparing the transmission for the Towards and Away delay lines, the difference between the conversion in port 1 and 2 was estimated. A transmission difference of 44~dB can be seen in Fig.~\ref{fig:RTD}c and Table~\ref{tab:samples} (\textit{Max D}). Half this difference defines the directivity of one FEUDT as the fraction between the power converted in the two acoustic ports. Samples FEUDT\_2-4 showed a maximum directivity of 22~dB for each FEUDT, which means that 99.4~$\%$ of the acoustic power went to port 1. Each FEUDT lost 0.6~\% of the power in the wrong direction, which can be compared to the theoretical IDT loss of 50~\%. Thus, the directionality accounts for \mbox{-0.06~dB} of the loss in the FEUDT delay line and \mbox{-6~dB} of the loss in IDT delay lines ($\gamma_\text{D}$ in Table~\ref{tab:samples}). There was an additional loss of about \mbox{-3.4~dB} in the Towards delay lines and about \mbox{-1.7~dB} in the IDT delay lines that cannot be attributed to directive loss. This power was lost either during electric/SAW conversion or during the propagation in the delay line. The directional loss accounts for more than 90~\% of the total loss in the IDT delay lines, whereas it accounts for only 3~\% percent of the total loss for the FEUDT delay lines. 

The conversion loss is due to imperfections in the transducers. At 10~mK the aluminum electrodes are superconducting, which eliminates resistive losses in the transducers. It has been shown, by using superconducting niobium FEUDTs at 3.5~K, that the electrode resistance has a much bigger effect on the insertion loss than other loss mechanisms \cite{Yamanouchi1998scNb}. 

The propagation losses for all samples are expected to be similar since they were fabricated simultaneously on the same wafer. In addition, transducer orientation and separation were kept fixed. This separation distance was assumed to be between the edge of the transducers, but the SAW travels further underneath it. Using the center of the transducers ($L+N_p\lambda$) as the reference point instead, the SAW propagates an additional 240 or 165~$\mu$m in the FEUDT delay lines and 54~$\mu$m further in the IDT delay lines (Table~\ref{tab:samples}). 

The losses during the SAW propagation may include beam steering, diffraction and viscous damping \cite{Slobodnik1970,Slobodnik1978}. The loss due to beam steering ($\gamma_\text{bs}$ in Table~\ref{tab:samples}) was estimated to about \mbox{-0.3~dB}, by assuming an alignment error of 0.1$^{\circ}$ and using the slope of the power flow angle \mbox{-1.083} for LiNbO$_3$ \cite{Slobodnik1970}. The diffraction loss ($\gamma_\text{diff}$) was linearly extrapolated from the results in \cite{Szabo1973} and was estimated to around \mbox{-0.7~dB}.  Damping from gas loading could be ignored, since the experiments were performed in vacuum. The the calculated loss due to viscous damping was a bit more than \mbox{-0.1~dB}, using a viscous damping factor of 0.88~dB/$\mu$s for room temperature \cite{Slobodnik1978}. Hence, the estimated diffraction is the dominant propagation loss. In total, the theoretically estimated propagation loss was around \mbox{-1.1~dB}. Consequently, we can account for \mbox{-1.3~dB} and \mbox{-7.1~dB} of the loss in the FEUDT and IDT delay lines, respectively. This leaves an unexplained loss ($\gamma_\text{ue}$ in Table~\ref{tab:samples}) of \mbox{-2.2$\pm$0.8~dB} in the FEUDT delay lines and \mbox{-0.5$\pm$0.2~dB}  in the IDT delay lines that cannot be attributed to directive nor propagation losses. The higher loss in the FEUDTs seems to be related to their higher number of unit cells. The loss per unit cell, \mbox{-0.007$\pm$0.003~dB}, was the same for both types of transducers. 

%\subsection{Power conservation from attenuation fits}
The losses were further investigated by comparing the fitted attenuation of each SAW transit in the delay lines. These fits used the same models as the main transmission. Each transit has undergone conversion into SAW, propagated at least one time in the delay line and then undergone conversion from SAW  back to electric signal. The difference between the transits is the number of times the SAW has been acoustically reflected and has propagated across the delay line. Thus, the sum of the loss due to propagation and acoustic reflection could be extracted. This loss was less than \mbox{-1.4~dB} for all samples. Subtracting the estimated value of the propagation loss, less than \mbox{-0.3~dB} of the signal was lost per acoustic reflection. 

If the same argument is used for the main transmission (two signal conversions and one transit), less than \mbox{-0.4~dB} was lost per IDT every time the signal was converted. Similarly, the maximum loss every time the signal was converted was \mbox{-0.9$\pm$0.6~dB} per FEUDT. This results in a maximum loss of \mbox{-0.01~dB} per unit cell for all transducers. The nature of this conversion loss is unknown, but it may include conversion to bulk waves. 
 
%\subsection{Bandwidth}
The COM theory indicates that FEUDTs need more than 100 unit cells and larger than 25 um electrode overlap in order to achieve a minimum 20 dB directivity and impedance matching to 50 Ohms. For comparison, IDTs with 35 um electrode overlap are impedance matched with only 36 unit cells. The number of unit cells reduces the bandwidth according to $0.9f_0/N_p$ \cite{Datta}, which explains the results in Table~\ref{tab:samples} where the FEUDT with 160 unit cells had a 5 times smaller bandwidth than the IDTs. Although the narrow bandwidth is useful for on-chip filtering, it can be a limitation in quantum SAW experiments. 

%\section{Quantum application discussion}
In quantum SAW experiments, the qubit uses an interdigital transducer to couple to SAW phonons. At the center frequency of the qubit transducer $f_q$, this coupling is given by $\Gamma_{10}/2\pi=0.5N_{q}K^2f_{q}$~\cite{MVG,bookchapter} where $N_q$ is the number of unit cells of the qubit transducer. Since these has to be at least one unit cell, the minimum coupling $\Gamma/2\pi$ is approximately 100~MHz for a qubit on LiNbO$_3$ at 2.3~GHz. If a pick up transducer has a smaller bandwidth than the qubit coupling, the phonons emitted from the qubit will not activate all unit cells in the pick up transducer. Thus, there is a trade-off between bandwidth and directivity that needs to be optimized for a given experiment. 

%\section{Conclusion and future}
In conclusion, we have demonstrated a directivity of 22~dB for superconducting floating electrode unidirectional transducers at 2.3~GHz and 10~mK on lithium niobate. The unidirectional delay lines have approximately 4.7~dB less insertion loss than standard interdigital transducers. The improved phonon to photon conversion of unidirectional transducer compared to ordinary interdigital transducer is useful for studying quantum physics with surface acoustic waves. The directivity and impedance matching of a floating electrode unidirectional transducer come at the expense of bandwidth. 

\begin{acknowledgments}
We wish to acknowledge financial support from The Knut and Alice Wallenberg Foundation and The Swedish Research Council. The samples were made at the Nanofabrication Laboratory at Chalmers University of Technology. We also acknowledge discussions with Martin V. Gustafsson, G\"oran Johansson, Anton Frisk-Kockum, Haruki Sanada, Andreas Josefsson Ask, Gustav Andersson and Baladitya Suri.
\end{acknowledgments}

\bibliography{bib_Article_aip}

\providecommand{\noopsort}[1]{}\providecommand{\singleletter}[1]{#1}%
\begin{thebibliography}{29}%
\makeatletter
\providecommand \@ifxundefined [1]{%
 \@ifx{#1\undefined}
}%
\providecommand \@ifnum [1]{%
 \ifnum #1\expandafter \@firstoftwo
 \else \expandafter \@secondoftwo
 \fi
}%
\providecommand \@ifx [1]{%
 \ifx #1\expandafter \@firstoftwo
 \else \expandafter \@secondoftwo
 \fi
}%
\providecommand \natexlab [1]{#1}%
\providecommand \enquote  [1]{``#1''}%
\providecommand \bibnamefont  [1]{#1}%
\providecommand \bibfnamefont [1]{#1}%
\providecommand \citenamefont [1]{#1}%
\providecommand \href@noop [0]{\@secondoftwo}%
\providecommand \href [0]{\begingroup \@sanitize@url \@href}%
\providecommand \@href[1]{\@@startlink{#1}\@@href}%
\providecommand \@@href[1]{\endgroup#1\@@endlink}%
\providecommand \@sanitize@url [0]{\catcode `\\12\catcode `\$12\catcode
  `\&12\catcode `\#12\catcode `\^12\catcode `\_12\catcode `\%12\relax}%
\providecommand \@@startlink[1]{}%
\providecommand \@@endlink[0]{}%
\providecommand \url  [0]{\begingroup\@sanitize@url \@url }%
\providecommand \@url [1]{\endgroup\@href {#1}{\urlprefix }}%
\providecommand \urlprefix  [0]{URL }%
\providecommand \Eprint [0]{\href }%
\providecommand \doibase [0]{http://dx.doi.org/}%
\providecommand \selectlanguage [0]{\@gobble}%
\providecommand \bibinfo  [0]{\@secondoftwo}%
\providecommand \bibfield  [0]{\@secondoftwo}%
\providecommand \translation [1]{[#1]}%
\providecommand \BibitemOpen [0]{}%
\providecommand \bibitemStop [0]{}%
\providecommand \bibitemNoStop [0]{.\EOS\space}%
\providecommand \EOS [0]{\spacefactor3000\relax}%
\providecommand \BibitemShut  [1]{\csname bibitem#1\endcsname}%
\let\auto@bib@innerbib\@empty
%</preamble>
\bibitem [{\citenamefont {Rayleigh}(1885)}]{Rayleigh1885}%
  \BibitemOpen
  \bibfield  {author} {\bibinfo {author} {\bibfnamefont {L.}~\bibnamefont
  {Rayleigh}},\ }\href@noop {} {\bibfield  {journal} {\bibinfo  {journal}
  {Proceedings of the London Mathematical Society}\ ,\ \bibinfo {pages} {4}}
  (\bibinfo {year} {1885})}\BibitemShut {NoStop}%
\bibitem [{\citenamefont {Gustafsson}\ \emph {et~al.}(2012)\citenamefont
  {Gustafsson}, \citenamefont {Santos}, \citenamefont {Johansson},\ and\
  \citenamefont {Delsing}}]{MVG2012}%
  \BibitemOpen
  \bibfield  {author} {\bibinfo {author} {\bibfnamefont {M.~V.}\ \bibnamefont
  {Gustafsson}}, \bibinfo {author} {\bibfnamefont {P.}~\bibnamefont {Santos}},
  \bibinfo {author} {\bibfnamefont {G.}~\bibnamefont {Johansson}}, \ and\
  \bibinfo {author} {\bibfnamefont {P.}~\bibnamefont {Delsing}},\ }\href@noop
  {} {\bibfield  {journal} {\bibinfo  {journal} {Nature Phys.}\ }\textbf
  {\bibinfo {volume} {8}},\ \bibinfo {pages} {338} (\bibinfo {year}
  {2012})}\BibitemShut {NoStop}%
\bibitem [{\citenamefont {Gustafsson}\ \emph {et~al.}(2014)\citenamefont
  {Gustafsson}, \citenamefont {Aref}, \citenamefont {Kockum}, \citenamefont
  {Ekstr{\"o}m}, \citenamefont {Johansson},\ and\ \citenamefont
  {Delsing}}]{MVG}%
  \BibitemOpen
  \bibfield  {author} {\bibinfo {author} {\bibfnamefont {M.~V.}\ \bibnamefont
  {Gustafsson}}, \bibinfo {author} {\bibfnamefont {T.}~\bibnamefont {Aref}},
  \bibinfo {author} {\bibfnamefont {A.~F.}\ \bibnamefont {Kockum}}, \bibinfo
  {author} {\bibfnamefont {M.~K.}\ \bibnamefont {Ekstr{\"o}m}}, \bibinfo
  {author} {\bibfnamefont {G.}~\bibnamefont {Johansson}}, \ and\ \bibinfo
  {author} {\bibfnamefont {P.}~\bibnamefont {Delsing}},\ }\href@noop {}
  {\bibfield  {journal} {\bibinfo  {journal} {Science}\ }\textbf {\bibinfo
  {volume} {346}},\ \bibinfo {pages} {207} (\bibinfo {year}
  {2014})}\BibitemShut {NoStop}%
\bibitem [{\citenamefont {Frisk~Kockum}\ \emph {et~al.}(2014)\citenamefont
  {Frisk~Kockum}, \citenamefont {Delsing},\ and\ \citenamefont
  {Johansson}}]{Kockum2014}%
  \BibitemOpen
  \bibfield  {author} {\bibinfo {author} {\bibfnamefont {A.}~\bibnamefont
  {Frisk~Kockum}}, \bibinfo {author} {\bibfnamefont {P.}~\bibnamefont
  {Delsing}}, \ and\ \bibinfo {author} {\bibfnamefont {G.}~\bibnamefont
  {Johansson}},\ }\href {\doibase 10.1103/PhysRevA.90.013837} {\bibfield
  {journal} {\bibinfo  {journal} {Phys. Rev. A}\ }\textbf {\bibinfo {volume}
  {90}},\ \bibinfo {pages} {013837} (\bibinfo {year} {2014})}\BibitemShut
  {NoStop}%
\bibitem [{\citenamefont {Manenti}\ \emph {et~al.}(2016)\citenamefont
  {Manenti}, \citenamefont {Peterer}, \citenamefont {Nersisyan}, \citenamefont
  {Magnusson}, \citenamefont {Patterson},\ and\ \citenamefont
  {Leek}}]{Manenti2016}%
  \BibitemOpen
  \bibfield  {author} {\bibinfo {author} {\bibfnamefont {R.}~\bibnamefont
  {Manenti}}, \bibinfo {author} {\bibfnamefont {M.~J.}\ \bibnamefont
  {Peterer}}, \bibinfo {author} {\bibfnamefont {A.}~\bibnamefont {Nersisyan}},
  \bibinfo {author} {\bibfnamefont {E.~B.}\ \bibnamefont {Magnusson}}, \bibinfo
  {author} {\bibfnamefont {A.}~\bibnamefont {Patterson}}, \ and\ \bibinfo
  {author} {\bibfnamefont {P.~J.}\ \bibnamefont {Leek}},\ }\href {\doibase
  10.1103/PhysRevB.93.041411} {\bibfield  {journal} {\bibinfo  {journal} {Phys.
  Rev. B}\ }\textbf {\bibinfo {volume} {93}},\ \bibinfo {pages} {041411}
  (\bibinfo {year} {2016})}\BibitemShut {NoStop}%
\bibitem [{\citenamefont {Magnusson}\ \emph {et~al.}(2015)\citenamefont
  {Magnusson}, \citenamefont {Williams}, \citenamefont {Manenti}, \citenamefont
  {Nam}, \citenamefont {Nersisyan}, \citenamefont {Peterer}, \citenamefont
  {Ardavan},\ and\ \citenamefont {Leek}}]{Magnusson2015}%
  \BibitemOpen
  \bibfield  {author} {\bibinfo {author} {\bibfnamefont {E.~B.}\ \bibnamefont
  {Magnusson}}, \bibinfo {author} {\bibfnamefont {B.~H.}\ \bibnamefont
  {Williams}}, \bibinfo {author} {\bibfnamefont {R.}~\bibnamefont {Manenti}},
  \bibinfo {author} {\bibfnamefont {M.}~\bibnamefont {Nam}}, \bibinfo {author}
  {\bibfnamefont {A.}~\bibnamefont {Nersisyan}}, \bibinfo {author}
  {\bibfnamefont {M.}~\bibnamefont {Peterer}}, \bibinfo {author} {\bibfnamefont
  {A.}~\bibnamefont {Ardavan}}, \ and\ \bibinfo {author} {\bibfnamefont
  {P.~J.}\ \bibnamefont {Leek}},\ }\href {\doibase
  http://dx.doi.org/10.1063/1.4908248} {\bibfield  {journal} {\bibinfo
  {journal} {App. Phys. Lett.}\ }\textbf {\bibinfo {volume} {106}},\ \bibinfo
  {pages} {063509} (\bibinfo {year} {2015})}\BibitemShut {NoStop}%
\bibitem [{\citenamefont {Naber}\ \emph {et~al.}(2006)\citenamefont {Naber},
  \citenamefont {Fujisawa}, \citenamefont {Liu},\ and\ \citenamefont {van~der
  Wiel}}]{Naber2006SAWDQD}%
  \BibitemOpen
  \bibfield  {author} {\bibinfo {author} {\bibfnamefont {W.~J.~M.}\
  \bibnamefont {Naber}}, \bibinfo {author} {\bibfnamefont {T.}~\bibnamefont
  {Fujisawa}}, \bibinfo {author} {\bibfnamefont {H.~W.}\ \bibnamefont {Liu}}, \
  and\ \bibinfo {author} {\bibfnamefont {W.~G.}\ \bibnamefont {van~der Wiel}},\
  }\href {\doibase 10.1103/PhysRevLett.96.136807} {\bibfield  {journal}
  {\bibinfo  {journal} {Phys. Rev. Lett.}\ }\textbf {\bibinfo {volume} {96}},\
  \bibinfo {pages} {136807} (\bibinfo {year} {2006})}\BibitemShut {NoStop}%
\bibitem [{\citenamefont {Barnes}\ \emph {et~al.}(2000)\citenamefont {Barnes},
  \citenamefont {Shilton},\ and\ \citenamefont {Robinson}}]{Barnes2000}%
  \BibitemOpen
  \bibfield  {author} {\bibinfo {author} {\bibfnamefont {C.~H.~W.}\
  \bibnamefont {Barnes}}, \bibinfo {author} {\bibfnamefont {J.~M.}\
  \bibnamefont {Shilton}}, \ and\ \bibinfo {author} {\bibfnamefont {A.~M.}\
  \bibnamefont {Robinson}},\ }\href {\doibase 10.1103/PhysRevB.62.8410}
  {\bibfield  {journal} {\bibinfo  {journal} {Phys. Rev. B}\ }\textbf {\bibinfo
  {volume} {62}},\ \bibinfo {pages} {8410} (\bibinfo {year}
  {2000})}\BibitemShut {NoStop}%
\bibitem [{\citenamefont {Hermelin}\ \emph {et~al.}(2011)\citenamefont
  {Hermelin}, \citenamefont {Takada}, \citenamefont {Yamamoto}, \citenamefont
  {Tarucha}, \citenamefont {Wieck}, \citenamefont {Saminadayar}, \citenamefont
  {B{\"a}uerle},\ and\ \citenamefont {Meunier}}]{Hermelin2011}%
  \BibitemOpen
  \bibfield  {author} {\bibinfo {author} {\bibfnamefont {S.}~\bibnamefont
  {Hermelin}}, \bibinfo {author} {\bibfnamefont {S.}~\bibnamefont {Takada}},
  \bibinfo {author} {\bibfnamefont {M.}~\bibnamefont {Yamamoto}}, \bibinfo
  {author} {\bibfnamefont {S.}~\bibnamefont {Tarucha}}, \bibinfo {author}
  {\bibfnamefont {A.~D.}\ \bibnamefont {Wieck}}, \bibinfo {author}
  {\bibfnamefont {L.}~\bibnamefont {Saminadayar}}, \bibinfo {author}
  {\bibfnamefont {C.}~\bibnamefont {B{\"a}uerle}}, \ and\ \bibinfo {author}
  {\bibfnamefont {T.}~\bibnamefont {Meunier}},\ }\href@noop {} {\bibfield
  {journal} {\bibinfo  {journal} {Nature}\ }\textbf {\bibinfo {volume} {477}},\
  \bibinfo {pages} {435} (\bibinfo {year} {2011})}\BibitemShut {NoStop}%
\bibitem [{\citenamefont {McNeil}\ \emph {et~al.}(2011)\citenamefont {McNeil},
  \citenamefont {Kataoka}, \citenamefont {Ford}, \citenamefont {Barnes},
  \citenamefont {Anderson}, \citenamefont {Jones}, \citenamefont {Farrer},\
  and\ \citenamefont {Ritchie}}]{McNeil2011}%
  \BibitemOpen
  \bibfield  {author} {\bibinfo {author} {\bibfnamefont {R.~P.~G.}\
  \bibnamefont {McNeil}}, \bibinfo {author} {\bibfnamefont {M.}~\bibnamefont
  {Kataoka}}, \bibinfo {author} {\bibfnamefont {C.~J.~B.}\ \bibnamefont
  {Ford}}, \bibinfo {author} {\bibfnamefont {C.~H.~W.}\ \bibnamefont {Barnes}},
  \bibinfo {author} {\bibfnamefont {D.}~\bibnamefont {Anderson}}, \bibinfo
  {author} {\bibfnamefont {G.~A.~C.}\ \bibnamefont {Jones}}, \bibinfo {author}
  {\bibfnamefont {I.}~\bibnamefont {Farrer}}, \ and\ \bibinfo {author}
  {\bibfnamefont {D.~A.}\ \bibnamefont {Ritchie}},\ }\href@noop {} {\bibfield
  {journal} {\bibinfo  {journal} {Nature}\ }\textbf {\bibinfo {volume} {477}},\
  \bibinfo {pages} {439} (\bibinfo {year} {2011})}\BibitemShut {NoStop}%
\bibitem [{\citenamefont {Cerda-M\'endez}\ \emph {et~al.}(2010)\citenamefont
  {Cerda-M\'endez}, \citenamefont {Krizhanovskii}, \citenamefont {Wouters},
  \citenamefont {Bradley}, \citenamefont {Biermann}, \citenamefont {Guda},
  \citenamefont {Hey}, \citenamefont {Santos}, \citenamefont {Sarkar},\ and\
  \citenamefont {Skolnick}}]{Santos2010}%
  \BibitemOpen
  \bibfield  {author} {\bibinfo {author} {\bibfnamefont {E.~A.}\ \bibnamefont
  {Cerda-M\'endez}}, \bibinfo {author} {\bibfnamefont {D.~N.}\ \bibnamefont
  {Krizhanovskii}}, \bibinfo {author} {\bibfnamefont {M.}~\bibnamefont
  {Wouters}}, \bibinfo {author} {\bibfnamefont {R.}~\bibnamefont {Bradley}},
  \bibinfo {author} {\bibfnamefont {K.}~\bibnamefont {Biermann}}, \bibinfo
  {author} {\bibfnamefont {K.}~\bibnamefont {Guda}}, \bibinfo {author}
  {\bibfnamefont {R.}~\bibnamefont {Hey}}, \bibinfo {author} {\bibfnamefont
  {P.~V.}\ \bibnamefont {Santos}}, \bibinfo {author} {\bibfnamefont
  {D.}~\bibnamefont {Sarkar}}, \ and\ \bibinfo {author} {\bibfnamefont {M.~S.}\
  \bibnamefont {Skolnick}},\ }\href {\doibase 10.1103/PhysRevLett.105.116402}
  {\bibfield  {journal} {\bibinfo  {journal} {Phys. Rev. Lett.}\ }\textbf
  {\bibinfo {volume} {105}},\ \bibinfo {pages} {116402} (\bibinfo {year}
  {2010})}\BibitemShut {NoStop}%
\bibitem [{\citenamefont {Aref}\ \emph {et~al.}(2016)\citenamefont {Aref},
  \citenamefont {Delsing}, \citenamefont {Ekstr{\"o}m}, \citenamefont {Kockum},
  \citenamefont {Gustafsson}, \citenamefont {Johansson}, \citenamefont {Leek},
  \citenamefont {Magnusson},\ and\ \citenamefont {Manenti}}]{bookchapter}%
  \BibitemOpen
  \bibfield  {author} {\bibinfo {author} {\bibfnamefont {T.}~\bibnamefont
  {Aref}}, \bibinfo {author} {\bibfnamefont {P.}~\bibnamefont {Delsing}},
  \bibinfo {author} {\bibfnamefont {M.~K.}\ \bibnamefont {Ekstr{\"o}m}},
  \bibinfo {author} {\bibfnamefont {A.~F.}\ \bibnamefont {Kockum}}, \bibinfo
  {author} {\bibfnamefont {M.~V.}\ \bibnamefont {Gustafsson}}, \bibinfo
  {author} {\bibfnamefont {G.}~\bibnamefont {Johansson}}, \bibinfo {author}
  {\bibfnamefont {P.~J.}\ \bibnamefont {Leek}}, \bibinfo {author}
  {\bibfnamefont {E.}~\bibnamefont {Magnusson}}, \ and\ \bibinfo {author}
  {\bibfnamefont {R.}~\bibnamefont {Manenti}},\ }\enquote {\bibinfo {title}
  {Quantum acoustics with surface acoustic waves},}\ in\ \href {\doibase
  10.1007/978-3-319-24091-6_9} {\emph {\bibinfo {booktitle} {Superconducting
  Devices in Quantum Optics}}},\ \bibinfo {editor} {edited by\ \bibinfo
  {editor} {\bibfnamefont {H.~R.}\ \bibnamefont {Hadfield}}\ and\ \bibinfo
  {editor} {\bibfnamefont {G.}~\bibnamefont {Johansson}}}\ (\bibinfo
  {publisher} {Springer International Publishing},\ \bibinfo {address} {Cham},\
  \bibinfo {year} {2016})\ pp.\ \bibinfo {pages} {217--244}\BibitemShut
  {NoStop}%
\bibitem [{\citenamefont {Collins}\ \emph {et~al.}(1969)\citenamefont
  {Collins}, \citenamefont {Gerard}, \citenamefont {Reeder},\ and\
  \citenamefont {Shaw}}]{Collins1969}%
  \BibitemOpen
  \bibfield  {author} {\bibinfo {author} {\bibfnamefont {J.~H.}\ \bibnamefont
  {Collins}}, \bibinfo {author} {\bibfnamefont {H.~M.}\ \bibnamefont {Gerard}},
  \bibinfo {author} {\bibfnamefont {T.~M.}\ \bibnamefont {Reeder}}, \ and\
  \bibinfo {author} {\bibfnamefont {H.~J.}\ \bibnamefont {Shaw}},\ }\href@noop
  {} {\bibfield  {journal} {\bibinfo  {journal} {Proc. IEEE}\ }\textbf
  {\bibinfo {volume} {57}},\ \bibinfo {pages} {833} (\bibinfo {year}
  {1969})}\BibitemShut {NoStop}%
\bibitem [{\citenamefont {Morgan}(2007)}]{Morgan}%
  \BibitemOpen
  \bibfield  {author} {\bibinfo {author} {\bibfnamefont {D.}~\bibnamefont
  {Morgan}},\ }\href@noop {} {\emph {\bibinfo {title} {Surface Acoustic Wave
  Filters}}},\ \bibinfo {edition} {2nd}\ ed.\ (\bibinfo  {publisher} {Academic
  Press},\ \bibinfo {address} {Amsterdam},\ \bibinfo {year} {2007})\BibitemShut
  {NoStop}%
\bibitem [{\citenamefont {Datta}(1986)}]{Datta}%
  \BibitemOpen
  \bibfield  {author} {\bibinfo {author} {\bibfnamefont {S.}~\bibnamefont
  {Datta}},\ }\href@noop {} {\emph {\bibinfo {title} {Surface Acoustic Wave
  Devices}}}\ (\bibinfo  {publisher} {Prentice-Hall},\ \bibinfo {address}
  {Englewood Cliffs, NJ},\ \bibinfo {year} {1986})\BibitemShut {NoStop}%
\bibitem [{\citenamefont {Yamanouchi}\ \emph {et~al.}(1992)\citenamefont
  {Yamanouchi}, \citenamefont {Lee}, \citenamefont {Yamamoto}, \citenamefont
  {Meguro},\ and\ \citenamefont {Odagawa}}]{Yamanouchi1992GHzFEUDT}%
  \BibitemOpen
  \bibfield  {author} {\bibinfo {author} {\bibfnamefont {K.}~\bibnamefont
  {Yamanouchi}}, \bibinfo {author} {\bibfnamefont {C.~H.~S.}\ \bibnamefont
  {Lee}}, \bibinfo {author} {\bibfnamefont {K.}~\bibnamefont {Yamamoto}},
  \bibinfo {author} {\bibfnamefont {T.}~\bibnamefont {Meguro}}, \ and\ \bibinfo
  {author} {\bibfnamefont {H.}~\bibnamefont {Odagawa}},\ }\href@noop {}
  {\bibfield  {journal} {\bibinfo  {journal} {IEEE Ultrasonics Symp.}\ ,\
  \bibinfo {pages} {139}} (\bibinfo {year} {1992})}\BibitemShut {NoStop}%
\bibitem [{\citenamefont {Yamanouchi}\ \emph {et~al.}(1994)\citenamefont
  {Yamanouchi}, \citenamefont {Meguro}, \citenamefont {Wagatsuma},
  \citenamefont {Odagawa},\ and\ \citenamefont
  {Yamamoto}}]{Yamanouchi1994highGHzFEUDT}%
  \BibitemOpen
  \bibfield  {author} {\bibinfo {author} {\bibfnamefont {K.}~\bibnamefont
  {Yamanouchi}}, \bibinfo {author} {\bibfnamefont {T.}~\bibnamefont {Meguro}},
  \bibinfo {author} {\bibfnamefont {Y.}~\bibnamefont {Wagatsuma}}, \bibinfo
  {author} {\bibfnamefont {H.}~\bibnamefont {Odagawa}}, \ and\ \bibinfo
  {author} {\bibfnamefont {K.}~\bibnamefont {Yamamoto}},\ }\href@noop {}
  {\bibfield  {journal} {\bibinfo  {journal} {J. of Appl. Phys.}\ }\textbf
  {\bibinfo {volume} {33}},\ \bibinfo {pages} {3018} (\bibinfo {year}
  {1994})}\BibitemShut {NoStop}%
\bibitem [{\citenamefont {Yamanouchi}(1998)}]{Yamanouchi1998scNb}%
  \BibitemOpen
  \bibfield  {author} {\bibinfo {author} {\bibfnamefont {K.}~\bibnamefont
  {Yamanouchi}},\ }\href@noop {} {\bibfield  {journal} {\bibinfo  {journal}
  {IEEE Ultrasonics Symp.}\ }\textbf {\bibinfo {volume} {1}},\ \bibinfo {pages}
  {57} (\bibinfo {year} {1998})}\BibitemShut {NoStop}%
\bibitem [{\citenamefont {Yamanouchi}\ and\ \citenamefont
  {Furuyashiki}(1984)}]{Yamanouchi1984}%
  \BibitemOpen
  \bibfield  {author} {\bibinfo {author} {\bibfnamefont {K.}~\bibnamefont
  {Yamanouchi}}\ and\ \bibinfo {author} {\bibfnamefont {H.}~\bibnamefont
  {Furuyashiki}},\ }\href@noop {} {\bibfield  {journal} {\bibinfo  {journal}
  {Electron. Lett.}\ }\textbf {\bibinfo {volume} {20}},\ \bibinfo {pages} {989}
  (\bibinfo {year} {1984})}\BibitemShut {NoStop}%
\bibitem [{\citenamefont {Bristol}\ \emph {et~al.}(1972)\citenamefont
  {Bristol}, \citenamefont {Jones}, \citenamefont {Snow},\ and\ \citenamefont
  {Smith}}]{Bristol1972}%
  \BibitemOpen
  \bibfield  {author} {\bibinfo {author} {\bibfnamefont {T.~W.}\ \bibnamefont
  {Bristol}}, \bibinfo {author} {\bibfnamefont {W.~R.}\ \bibnamefont {Jones}},
  \bibinfo {author} {\bibfnamefont {P.~B.}\ \bibnamefont {Snow}}, \ and\
  \bibinfo {author} {\bibfnamefont {W.~R.}\ \bibnamefont {Smith}},\ }in\ \href
  {\doibase 10.1109/ULTSYM.1972.196097} {\emph {\bibinfo {booktitle} {1972
  Ultrasonics Symposium}}}\ (\bibinfo  {publisher} {IEEE},\ \bibinfo {year}
  {1972})\ pp.\ \bibinfo {pages} {343--345}\BibitemShut {NoStop}%
\bibitem [{\citenamefont {Smith}\ \emph {et~al.}(1969)\citenamefont {Smith},
  \citenamefont {Gerard}, \citenamefont {Collins}, \citenamefont {Reeder},\
  and\ \citenamefont {Shaw}}]{Smith1969}%
  \BibitemOpen
  \bibfield  {author} {\bibinfo {author} {\bibfnamefont {W.~R.}\ \bibnamefont
  {Smith}}, \bibinfo {author} {\bibfnamefont {H.~M.}\ \bibnamefont {Gerard}},
  \bibinfo {author} {\bibfnamefont {J.~H.}\ \bibnamefont {Collins}}, \bibinfo
  {author} {\bibfnamefont {T.~M.}\ \bibnamefont {Reeder}}, \ and\ \bibinfo
  {author} {\bibfnamefont {H.~J.}\ \bibnamefont {Shaw}},\ }\href@noop {}
  {\bibfield  {journal} {\bibinfo  {journal} {IEEE Trans. Microwave Theory
  Tech.}\ }\textbf {\bibinfo {volume} {MTT-17}},\ \bibinfo {pages} {856}
  (\bibinfo {year} {1969})}\BibitemShut {NoStop}%
\bibitem [{\citenamefont {Takeuchi}\ and\ \citenamefont
  {Yamanouchi}(1993)}]{Takeuchi1993}%
  \BibitemOpen
  \bibfield  {author} {\bibinfo {author} {\bibfnamefont {M.}~\bibnamefont
  {Takeuchi}}\ and\ \bibinfo {author} {\bibfnamefont {K.}~\bibnamefont
  {Yamanouchi}},\ }\href@noop {} {\bibfield  {journal} {\bibinfo  {journal}
  {IEEE Trans. UFFC}\ }\textbf {\bibinfo {volume} {40}},\ \bibinfo {pages}
  {648} (\bibinfo {year} {1993})}\BibitemShut {NoStop}%
\bibitem [{\citenamefont {Morgan}(1999)}]{Morgan1999}%
  \BibitemOpen
  \bibfield  {author} {\bibinfo {author} {\bibfnamefont {D.~P.}\ \bibnamefont
  {Morgan}},\ }\href@noop {} {\bibfield  {journal} {\bibinfo  {journal} {IEEE
  Ultrasonics Symp.}\ }\textbf {\bibinfo {volume} {48}},\ \bibinfo {pages}
  {107} (\bibinfo {year} {1999})}\BibitemShut {NoStop}%
\bibitem [{\citenamefont {Morgan}(2000)}]{Morgan2000}%
  \BibitemOpen
  \bibfield  {author} {\bibinfo {author} {\bibfnamefont {D.~P.}\ \bibnamefont
  {Morgan}},\ }\href@noop {} {\bibfield  {journal} {\bibinfo  {journal} {IEEE
  Ultrasonics Symp.}\ }\textbf {\bibinfo {volume} {1}},\ \bibinfo {pages} {15}
  (\bibinfo {year} {2000})}\BibitemShut {NoStop}%
\bibitem [{\citenamefont {Morgan}(2001)}]{Morgan2001}%
  \BibitemOpen
  \bibfield  {author} {\bibinfo {author} {\bibfnamefont {D.~P.}\ \bibnamefont
  {Morgan}},\ }\href@noop {} {\bibfield  {journal} {\bibinfo  {journal} {IEEE
  Trans. UFFC}\ ,\ \bibinfo {pages} {15}} (\bibinfo {year} {2001})}\BibitemShut
  {NoStop}%
\bibitem [{\citenamefont {Takeuchi}\ and\ \citenamefont
  {Yamanouchi}(1988)}]{Takeuchi1988}%
  \BibitemOpen
  \bibfield  {author} {\bibinfo {author} {\bibfnamefont {M.}~\bibnamefont
  {Takeuchi}}\ and\ \bibinfo {author} {\bibfnamefont {K.}~\bibnamefont
  {Yamanouchi}},\ }\href@noop {} {\bibfield  {journal} {\bibinfo  {journal}
  {IEEE Ultrasonics Symp.}\ ,\ \bibinfo {pages} {57}} (\bibinfo {year}
  {1988})}\BibitemShut {NoStop}%
\bibitem [{\citenamefont {{Slobodnik Jr.}}\ \emph {et~al.}(1970)\citenamefont
  {{Slobodnik Jr.}}, \citenamefont {Carr1},\ and\ \citenamefont
  {Budreau}}]{Slobodnik1970}%
  \BibitemOpen
  \bibfield  {author} {\bibinfo {author} {\bibfnamefont {A.~J.}\ \bibnamefont
  {{Slobodnik Jr.}}}, \bibinfo {author} {\bibfnamefont {P.~H.}\ \bibnamefont
  {Carr1}}, \ and\ \bibinfo {author} {\bibfnamefont {A.~J.}\ \bibnamefont
  {Budreau}},\ }\href@noop {} {\bibfield  {journal} {\bibinfo  {journal} {J. of
  Appl. Phys.}\ }\textbf {\bibinfo {volume} {41}},\ \bibinfo {pages} {4380}
  (\bibinfo {year} {1970})}\BibitemShut {NoStop}%
\bibitem [{\citenamefont {{Slobodnik Jr.}}(1978)}]{Slobodnik1978}%
  \BibitemOpen
  \bibfield  {author} {\bibinfo {author} {\bibfnamefont {A.~J.}\ \bibnamefont
  {{Slobodnik Jr.}}},\ }\enquote {\bibinfo {title} {Materials and their
  influence on performance},}\ in\ \href {\doibase 10.1007/3-540-08575-0_13}
  {\emph {\bibinfo {booktitle} {Acoustic Surface Waves}}}\ (\bibinfo
  {publisher} {Springer Berlin Heidelberg},\ \bibinfo {address} {Berlin,
  Heidelberg},\ \bibinfo {year} {1978})\ pp.\ \bibinfo {pages}
  {225--303}\BibitemShut {NoStop}%
\bibitem [{\citenamefont {Szabo}\ and\ \citenamefont {Jr}(1973)}]{Szabo1973}%
  \BibitemOpen
  \bibfield  {author} {\bibinfo {author} {\bibfnamefont {T.}~\bibnamefont
  {Szabo}}\ and\ \bibinfo {author} {\bibfnamefont {A.~J.~S.}\ \bibnamefont
  {Jr}},\ }\href@noop {} {\bibfield  {journal} {\bibinfo  {journal} {IEEE
  Trans. Sonics Ultrason.}\ }\textbf {\bibinfo {volume} {SU-20}},\ \bibinfo
  {pages} {240} (\bibinfo {year} {1973})}\BibitemShut {NoStop}%
\end{thebibliography}%
\end{document}